\documentclass[sigconf]{acmart}
\AtBeginDocument{%
  \providecommand\BibTeX{{%
    \normalfont B\kern-0.5em{\scshape i\kern-0.25em b}\kern-0.8em\TeX}}}

\copyrightyear{2023} 
\acmYear{2023} 
\setcopyright{rightsretained} 
\acmConference[CHI EA '23]{Extended Abstracts of the 2023 CHI Conference on Human Factors in Computing Systems}{April 23--28, 2023}{Hamburg, Germany}
\acmBooktitle{Extended Abstracts of the 2023 CHI Conference on Human Factors in Computing Systems (CHI EA '23), April 23--28, 2023, Hamburg, Germany}\acmDOI{10.1145/3544549.3585869}
\acmISBN{978-1-4503-9422-2/23/04}

\usepackage{balance}
\usepackage{caption}
\usepackage{subcaption}
\usepackage{appendix}
\usepackage[ruled, lined, linesnumbered, commentsnumbered, longend]{algorithm2e}

\begin{document}

\title[Predicting Future Affective Reactions in Human-Computer Dialogue]{I Know Your Feelings Before You Do: Predicting Future Affective Reactions in Human-Computer Dialogue}


\author{Yuanchao Li}
\authornote{Part of this work was done while working on the JST ERATO ISHIGURO Symbiotic Human-Robot Interaction Project at Kawahara Lab at Kyoto University.}
\email{yuanchao.li@ed.ac.uk}
\affiliation{%
  \institution{University of Edinburgh}
  \country{UK}
}

\author{Koji Inoue}
\authornote{Authors contributed equally and are listed in alphabetical order.}
\email{inoue@sap.ist.i.kyoto-u.ac.jp}
\affiliation{%
  \institution{Kyoto University}
  \country{Japan}
}

\author{Leimin Tian}
\authornotemark[2]
\email{leimin.tian@monash.edu}
\affiliation{%
  \institution{Monash University}
  \country{Australia}
}

\author{Changzeng Fu}
\email{changzeng.fu@irl.sys.es.osaka-u.ac.jp}
\affiliation{%
  \institution{Osaka University \& RIKEN}
   \country{Japan}
  }

\author{Carlos Ishi}
\email{carlos@atr.jp}
\affiliation{%
  \institution{ATR \& RIKEN}
   \country{Japan}
}

\author{Hiroshi Ishiguro}
\email{ishiguro @irl.sys.es.osaka-u.ac.jp}
\affiliation{%
  \institution{Osaka University \& ATR}
   \country{Japan}
}

\author{Tatsuya Kawahara}
\email{kawahara@i.kyoto-u.ac.jp}
\affiliation{%
  \institution{Kyoto University}
  \country{Japan}
}

\author{Catherine Lai}
\email{c.lai@ed.ac.uk}
\affiliation{%
  \institution{University of Edinburgh}
  \country{UK}
}

\renewcommand{\shortauthors}{Li et al.}

\begin{abstract}
Current Spoken Dialogue Systems (SDSs) often serve as passive listeners that respond only after receiving user speech. To achieve human-like dialogue, we propose a novel future prediction architecture that allows an SDS to anticipate future affective reactions based on its current behaviors before the user speaks. In this work, we investigate two scenarios: speech and laughter. In speech, we propose to predict the user's future emotion based on its temporal relationship with the system's current emotion and its causal relationship with the system's current Dialogue Act (DA). In laughter, we propose to predict the occurrence and type of the user's laughter using the system's laughter behaviors in the current turn. Preliminary analysis of human-robot dialogue demonstrated synchronicity in the emotions and laughter displayed by the human and robot, as well as DA-emotion causality in their dialogue. This verifies that our architecture can contribute to the development of an anticipatory SDS.
\end{abstract}

\ccsdesc[500]{Human-centered computing~Human computer interaction (HCI)}

\keywords{emotion, dialogue act, interaction, laughter, spoken dialogue system}



\maketitle

\section{Introduction}
Spoken dialogue is a common and natural form of communication in human social interaction. Thus, we are witnessing a growing interest in advancing Spoken Dialogue Systems (SDSs) capable of delivering task-specific services, both in research and in commercial applications. For example, voice assistants, such as Amazon Echo or Google Home, are widely used for information queries in people's daily lives. Meanwhile, embodied SDSs, for instance the humanoid robot Pepper, have been deployed to enhance human workforce in application scenarios such as hospitality and elderly care. The majority of these SDSs, however, converse passively and utter words more as a matter of response than asking questions or leading the conversation on their own initiative. Furthermore, many existing SDSs are built on top of natural language processing and generation models developed with written text data, overlooking the rich conversational and affective phenomena unique to spoken dialogue, such as non-verbal vocalizations and affective bursts. As a consequence, current SDSs are often perceived as stagnant and mechanical. To mitigate this issue, researchers have been investigating various dialogue-specific behaviors, such as turn-taking, backchanneling, and laughter, as they have been found to serve important functions in conversation, including the marking of prominence, syntactic disambiguation, attitudinal reactions, uncertainty, and topic shifts \cite{ward2019prosodic,ward1996using,lala2019analysis,mazzocconi2020s}. 

Studies on these dialogue behaviors usually include both detection and synthesis tasks. The detection task aims at predicting user behaviors from the received signals, while the synthesis task generates system behaviors. For the detection task, acoustic features such as Mel Frequency Cepstral Coefficients (MFCCs) and prosodic features such as pitch, energy, and duration of pause-bounded phrases are typically used as cues for the detection of these human-like behaviors \cite{skantze2021turn,noguchi1998prosody,kaushik2015laughter}. With the recent advance of deep learning techniques, such detection has started to become more robust and is being applied in real-world applications \cite{kawahara2016prediction}. Compared to the detection task, the synthesis task is more challenging for two major reasons: 1) It depends on the accuracy of the detection task. If the user behaviors are classified incorrectly at the very beginning, the process for system behavior synthesis could become totally meaningless or even harmful to user engagement. For example, if the system detects turn-yielding cues but the user is actually holding the turn, then the user's speech will be interrupted by the system. 2) Unlike the detection task, where audio alone can achieve reasonable performance (although lexical cues often help), the synthesis task requires suitable generation of both acoustic and lexical behaviors. When synthesizing fillers and backchannels, the meanings largely depend on their morphological forms \cite{kawahara2016prediction,nakanishi2019generating,li2019expressing,li2022robotic}. To address this challenge, the majority of the synthesis task is still performed following a rule-based method. Take backchanneling as an example: First, the user's speech is converted into a sequence of words by Automatic Speech Recognition (ASR). Next, the focus word of the sequence is extracted. If the focus word matches any entries in the pre-built query-response database, the system generates a backchannel based on the query-response pair. Otherwise, the system generates a short backchannel, such as ``Yeah'' to indicate it is listening \cite{li2019expressing}.

Such a detection-rule-synthesis process is a widely adopted operation in SDSs, yet it has several limitations. First, this can lead to delayed responses due to the time taken (often correlates to the duration of the input speech) to process the user's speech and synthesize suitable responses. Such delay can accumulate when there are several detection components (e.g., dialogue act recognition, emotion recognition, and turn-taking detection). Second, previous research in linguistics and communication theories suggests that human listeners have the ability to anticipate the interlocutor's behavior in real time based on the dialogue context and history \cite{deksne2021predicting,poria2019emotion,tanaka2019dialogue}, and such predictive power is core to human brains \cite{arnal2012cortical,nagai2015predictive,philippsen2018understanding}. Furthermore, human listeners can start planning their responses or even talking before the interlocutor finishes, resulting in cooperative overlaps or appropriate interruptions that are key to establishing rapport and sympathy \cite{truong2007automatic,george2019should}. Current SDSs, however, are incapable of exhibiting such anticipatory and collaborative dialogue behaviors.

To alleviate this problem, recent SDS research has started to investigate the feasibility of enabling the system to actively lead the conversation instead of behaving as a passive follower. \citet{wu2019proactive} proposed a knowledge graph that sequentially changed the discussion topics following a given conversation goal to keep the dialogue as natural and engaging as possible. Besides, \citet{lala2017attentive} and \citet{li2022alzheimer} proposed attentive and proactive listening systems. These systems have a proactive initiator that can make the dialogue systems behave somewhat actively to ask a follow-up question related to the most recent topic or start a new topic. Moreover, proactive behaviors have proven helpful in rendering a more competent and reliable system that could ultimately lead to a more trustworthy interaction partner \cite{kraus2020effects}. Nevertheless, these functions dealt with only the linguistic aspect (e.g., spoken content) without considering the paralinguistic aspect (e.g., affective expressions).

Therefore, we are motivated to build an anticipatory SDS by endowing it with the ability to predict the future affective reactions of the user. Inspired by findings in cognitive science that humans can anticipate certain future events, including affective ones \cite{miceli2014expectancy,davidson2001toward,castelfranchi2011anticipation}, we propose an architecture that allows the SDS to mimic this human ability to predict affective reactions in the user's next turn based on its current turn. We consider two scenarios: speech and laughter, which are distinguished as two acoustic events (though co-occurrence also exists) \cite{truong2007automatic,reithinger1996predicting}. In the speech scenario, we look at the prediction of future emotions (valence and arousal). In the laughter scenario, we investigate the prediction of future laughter (occurrence and type). Moreover, we propose a self-correction and adaptation function that updates the future prediction model using the outputs of a recognition model on the user's speech. When the future prediction model has low confidence in its outputs, the recognition model generates outputs using the user's speech collected so far as ground truths to correct the prediction model. We conducted a preliminary analysis on human-robot dialogue, which confirms the feasibility of implementing the proposed anticipatory and adaptive architecture.

\section{Related Work}
A dialogue is made up of a series of utterances, with the previous response determined by the history information \cite{reithinger1996predicting,park2012already}. Previous studies have found that the emotions of the interlocutors have a mapping relationship in human-human conversation. In persuasion dialogue, \citet{acosta2011achieving} discovered that the listener's dimensional emotion (valence, arousal, dominance) can be predicted from the emotion expressed in the immediately preceding speaker's utterance. \citet{majumder2019dialoguernn} demonstrated that in dyadic conversation, the listener's discrete emotion can be predicted by the context given by the preceding utterances and the emotion expressed. Such relationships have been considered when designing SDSs: by having a virtual or embodied agent mirror a user's emotions, i.e., the ``affective mirror'' \cite{picard2000affective}.

Furthermore, emotion is affected by other aspects of dialogue, such as the Dialogue Act (DA), which represents the communicative function of an utterance. Despite the fact that there is a mutual influence between emotion and DA in conversations \cite{craggs2003annotating}, understanding of such relationships is limited. However, a recent study has found that there is a clear temporal causal relationship between DAs and emotions, providing specific emotion-DA and DA-emotion pairs. For instance, \textit{Happiness} of the speaker's utterance has a great chance of causing the DA of \textit{appreciation} in the following response. The DA of \textit{signal-non-understanding} and \textit{backchannel-question-form} usually raise \textit{surprise} \cite{cao2021causal}. Apart from emotions expressed through speech, affective bursts, especially laughter, are paralinguistic events that occur frequently in spontaneous dialogues \cite{truong2007automatic,melder2007affective}. Laughter usually shows a contagious phenomenon that hearing laughter from others is known to trigger laughter in ourselves \cite{provine1992contagious}. In dialogues, such a contagion is called ``shared laughter'' \cite{estow2007self,navarretta2016mirroring}. A recent work has developed a shared laughter system that can decide whether to generate social laughter, mirthful laughter, or not laugh, depending on the detection of user laughter \cite{inoue2022can}.

Based on these novel findings and developments, we can expect to advance SDSs by applying the causal relationships between emotion and DA, as well as the laughter mapping relationship, i.e., predicting the affective reactions of the upcoming user utterance to prepare its next response towards realizing proactive and affective behaviors. To the best of our knowledge, this is the first work to propose such an anticipatory SDS framework, which has the potential to lead to human-like affective dialogue capabilities in SDSs.



\section{Proposed Architecture for An Anticipatory SDS}
\subsection{The Speech Scenario}
\begin{figure}[!ht]
\centering
  \begin{subfigure}{.46\textwidth}
      \centering
      \includegraphics[width=\textwidth]{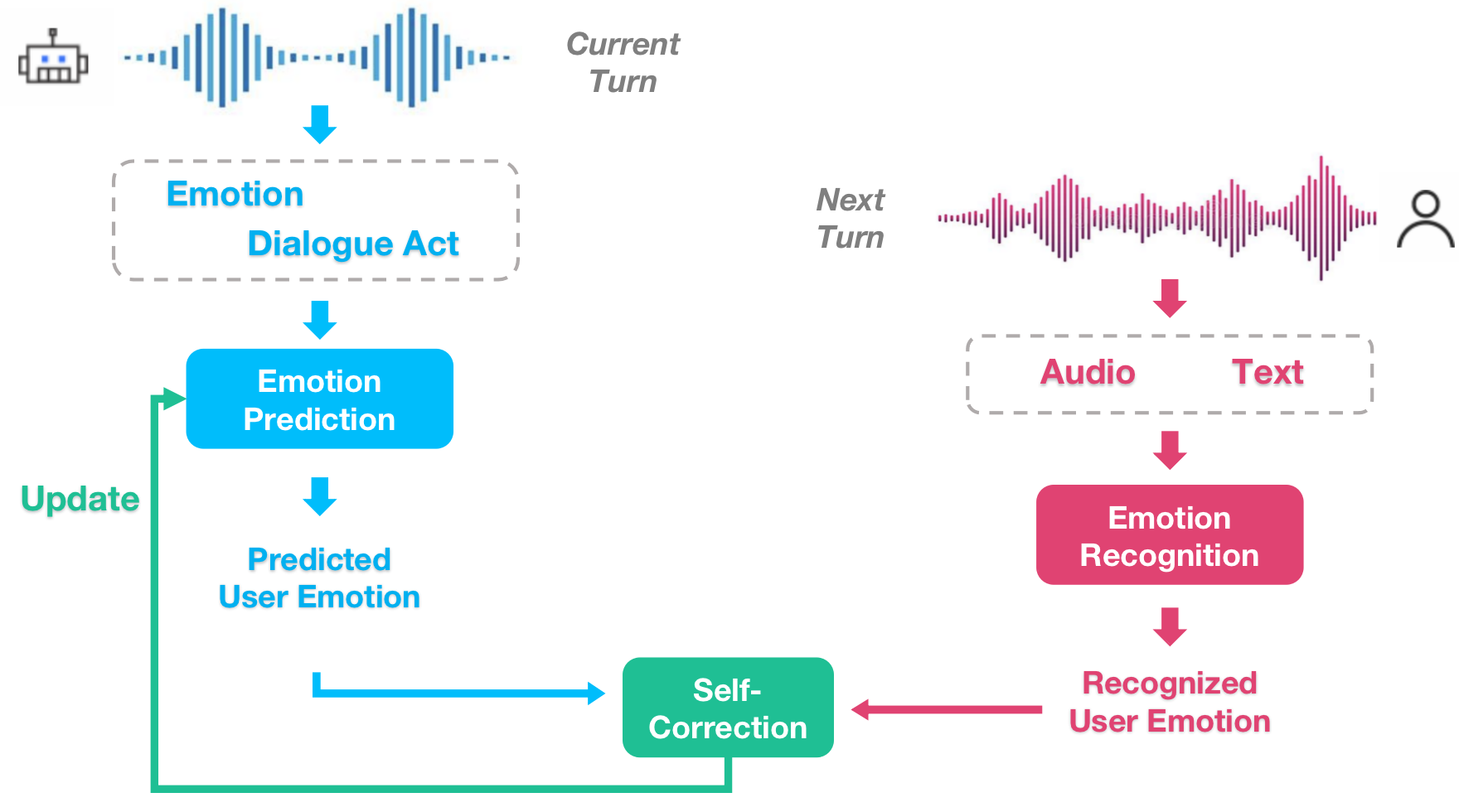}
      \caption{The Speech Scenario.}
      \label{speech}
      \end{subfigure}

  \vspace{5pt}    
  \hfill
  \begin{subfigure}{.46\textwidth}
     \centering
      \includegraphics[width=\textwidth]{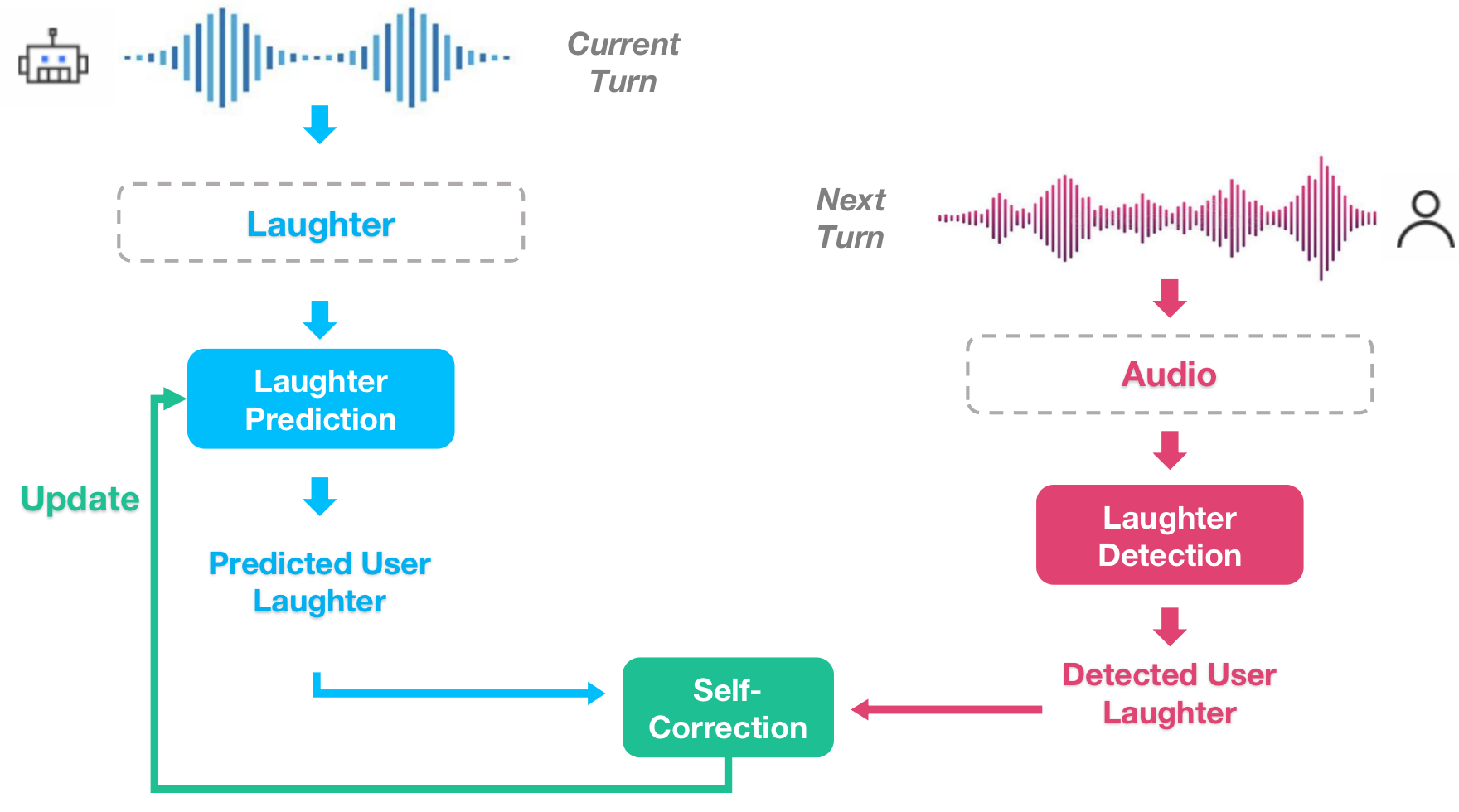}
      \caption{The Laughter Scenario.}
      \label{laughter}
     \end{subfigure}
  \caption{The proposed architecture for anticipatory dialogue.}
  \label{archi}
\end{figure}

In the speech scenario, as shown in Fig.~\ref{speech} and Algorithm~\ref{alg-speech}, there are three major components in the proposed architecture: \textit{emotion prediction}, \textit{emotion recognition}, and \textit{self-correction}. The emotion prediction model works as a future prediction function by taking the emotion and DA of the system's current turn as input and outputting an emotion as the prediction of the user's emotion in the next turn. When the prediction probability is high (i.e., the system is confident about its prediction), the system will take it as the user's future emotion and use it to help plan its next turn before the user speaks. Otherwise, the emotion recognition model starts to work by detecting the user's emotion once they finish speaking. The recognition result will be taken as ground truth to fine-tune the emotion prediction model if it has low confidence in its outputs and update the parameters of the prediction model via the self-correction function.

The emotion prediction model can be built by drawing on prior knowledge of emotion-emotion mapping \cite{li2019expressing} and the DA-emotion causal relationship \cite{cao2021causal}.
We conducted a preliminary analysis demonstrating that in human-robot dialogue, the users indeed mimic the robot's emotion, but the correlations show different patterns in the arousal and valence dimensions (described in Sec.~\ref{subsec:emo-pred-analysis}).
Besides, there is limited work on the causal relationship between DAs and emotions. Thus, we aim to expand this research to formulate the emotion prediction problem as a logistic regression model:
\begin{align}
    EM_{prd} = LR(EM_{cur} ,\ DA_{cur})
\end{align}
where the $EM_{prd}$ is the predicted user emotion in the next turn, and the $EM_{cur}$ and $DA_{cur}$ are the system's emotion and DA in the current turn, respectively. The regression model can be pre-trained on suitable corpora before being incorporated in the proposed architecture.

\begin{algorithm}[!ht]
    \small
    \SetKwFunction{isOddNumber}{isOddNumber}
    \SetKwInOut{KwIn}{Input}

    \KwIn{System's current emotion $EM_{cur}$ and dialogue act $DA_{cur}$;
    Predicted user emotion $EM_{prd}$; Recognized user emotion $EM_{rec}$; Pre-defined probability threshold $P_{thr1}$}
    
    \textbf{repeat} \\
    
    \For{system utterance}
    {
        {
        Generate $EM_{prd}$ from $EM_{cur}$ and $DA_{cur}$
        } \\
        \eIf{Probability of \  ${EM_{prd}} \geq P_{thr1}$}
        {
            Action \tcp*[f]{E.g., planning the system's next turn.}
        }
        {
            Generate $EM_{rec}$ \\
            Update the emotion prediction model  \\
        }
    }
    \textbf{until} no \textit{system utterance}
    \caption{The speech scenario}
    \label{alg-speech}
\end{algorithm}

\begin{algorithm}[!ht]
    \small
    \SetKwFunction{isOddNumber}{isOddNumber}
    \SetKwInOut{KwIn}{Input}

    \KwIn{System's current laughter $LA_{cur}$;
    Predicted user laughter $LA_{prd}$; Recognized user laughter $LA_{rec}$; Pre-defined probability threshold $P_{thr2}$}
    
    \textbf{repeat} \\
    \For{system laughter}
    {
        {
        Generate $LA_{prd}$ from $LA_{cur}$
        } \\
        \eIf{Probability of \  $LA_{prd} \geq P_{thr2}$}
        {
            Action \tcp*[f]{E.g., planning the system's next turn.}
        }
        {
            Generate $LA_{rec}$ \\
            Update the laughter prediction model  \\
        }
    }
    \textbf{until} no \textit{system laughter}
    \caption{The laughter scenario}
    \label{alg-laugh}
\end{algorithm}

Compared to emotion prediction, emotion recognition has been well studied, but the majority of these works did not take into consideration the real-life environment (e.g., noise), which is a long-standing problem for SDSs \cite{li2018towards}. Hence, we propose to incorporate text as additional information to tackle this problem. The text features are extracted from transcripts generated by the ASR component in SDSs, so the emotion recognition model needs to be robust to ASR errors. In a recent study, we developed a hierarchical cross-attention fusion model using both audio and text features for ASR error-robust emotion recognition \cite{li2022fusing}, which can be adopted in the proposed architecture. When using ASR transcripts, this model performed similarly to when using ground-truth transcripts.

The self-correction component follows the rule that when the confidence (i.e., probability) of the emotion prediction result is low, it starts to work by updating (i.e., fine-tuning) the prediction model using the outputs from the emotion recognition model as ground truths. Such a setting allows the emotion prediction component to dynamically adapt to the emotional expression habits of the human participants as the dialogue progresses. Like our human ability to make predictions, the longer the dialogue goes on, the more accurate the prediction becomes.

\subsection{The Laughter Scenario}
Similar to the speech scenario, there are three major components in the laughter scenario, as shown in Fig.~\ref{laughter} and Algorithm~\ref{alg-laugh}: \textit{laughter prediction}, \textit{laughter detection}, and \textit{self-correction}. Based on the system's laughter behavior in the current turn and the shared laughter relationship \cite{inoue2022can}, the laughter prediction model predicts the occurrence and type of the user's laughter in the next turn. If the prediction probability is low, the laughter detection will work by detecting the type of laughter from the user's response and updating the laughter prediction model via the self-correction function.

The laughter prediction model can be built by drawing on recent work that detects the occurrence and type of the user's laughter to generate the system's laughter \cite{inoue2022can}. We can use this finding reversely by predicting the occurrence and type of the user's laughter based on the system's laughter, which is easy to manipulate in SDSs. The laughter detection model takes acoustic features (e.g., MFCCs) and prosodic features (e.g., pitch and power) as input and feeds them to a stacked recurrent neural network. The recurrent neural network will be implemented using the bi-directional gated recurrent unit, whose feed-forward processing can work in real time, which is essential for SDSs. The self-correction follows the same idea as the speech scenario by updating (fine-tuning) the prediction model when its prediction probability is low.

\section{Preliminary Analysis}
\subsection{Corpora Description}

Although there are existing emotional dialogue corpora, most of them are not suitable for the purpose of this work. For example, IEMOCAP \cite{busso2008iemocap} contains spontaneous dialogue sessions, yet the improvisation is limited to a set of hypothetical scenarios, which is different from open domain natural dialogue. SEMAINE \cite{mckeown2010semaine} collected human-agent emotional dialogues, but the human users were not permitted to ask questions. The persuasive dialogue corpus used in \cite{acosta2011achieving} is not publicly available. Besides, none of the existing corpora contain sufficient occurrences and variations of laughter.

Therefore, we used two corpora from the JST ERATO ISHIGURO Symbiotic Human-Robot Interaction Project\footnote{\url{https://www.jst.go.jp/erato/ishiguro/en/index.html}} that contain spontaneous dialogue and rich laughter. The corpus for the speech scenario consists of spontaneous dialogue between human participants and a teleoperated humanoid robot ERICA \cite{glas2016erica}. During data collection, ERICA was teleoperated by a human operator in a Wizard of Oz (WoZ) manner. The participants were students ranging from 18 to 22 years old. Each dialogue session contained two phases and lasted around 15 minutes, and six sessions were conducted. In the first phase, ERICA introduced herself and talked with the participants about their lives, hobbies, and future plans. In the second phase, they talked about robots, especially about ERICA herself. During the dialogue, the robot led the dialogue, and the participant acted as a ``follower'', which is the scenario we hope to apply our proposed architecture to. The emotions are annotated as \textit{Valence}: -3 (extremely negative) to +3 (extremely positive), and \textit{Arousal}: -3 (extremely passive) to +3 (extremely active).

The corpus for the laughter scenario was collected under almost identical conditions, except that the aim was to get the teleoperators and the participants to know each other quickly \cite{inoue2016talking}. As a result, they behaved friendly by laughing frequently during such speed dating. Each dialogue lasts 10 to 15 minutes, and 82 dialogue sessions were conducted. The laughter was annotated as \textit{Social laughter}, \textit{Mirthful laughter}, and \textit{No laughter}.


\subsection{Exploring the Feasibility of Emotion Prediction}
\label{subsec:emo-pred-analysis}

To explore the relationship between the human participant's and the robot's emotions in dialogue, we analyzed the human-robot dialogue sessions from the first corpus. Our preliminary analysis found similar patterns in all six sessions. Because different sessions have different durations and different numbers of utterances, we could not average the emotion labels over the six sessions. Thus, we report one session containing 123 utterance pairs as an illustrative example to discuss our findings.

\begin{figure}
\centering
\includegraphics[width=0.47\textwidth]{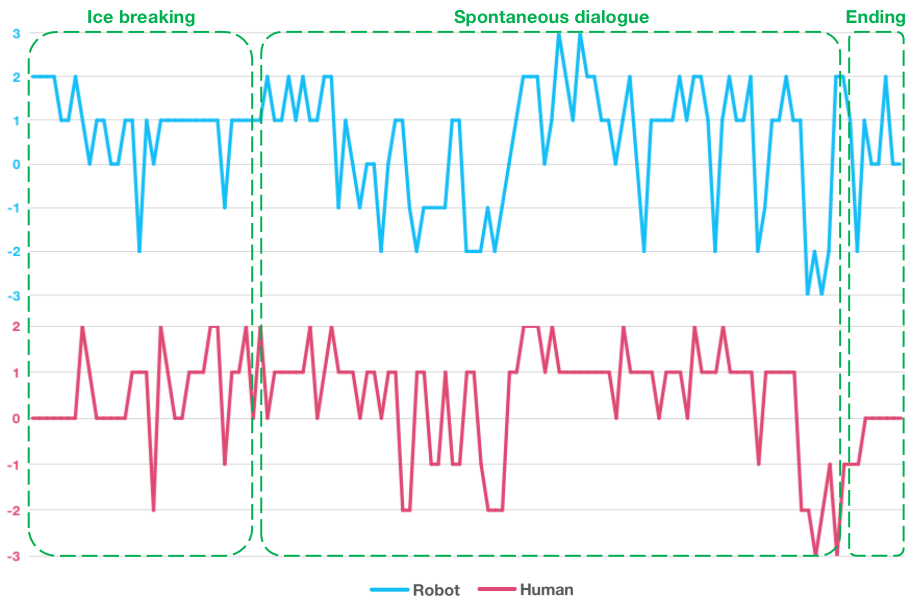}
\caption{Valence of the robot's and human's speech during an example dialogue session shows a mimicry pattern.}
\label{mimicry}
\end{figure}

\begin{table*}[!ht]
\caption{Annotated excerpt for analyzing the influence of dialogue acts on emotions in dialogue.}
\scalebox{0.91}{
\begin{tabular}{llccc}
\hline
\textbf{Speaker} & \textbf{Transcript}       & \textbf{Valence} & \textbf{Arousal} & \textbf{DA}              \\ \hline
\textit{Robot}         & Where are you from?       & +1               & +1               & Wh-question              \\
\textit{Participant}      & Tokushima, in Shikoku.    & +1               & +2               & Statement                \\
\textit{Robot}         & Fukushima?                & +2               & +2               & Signal-non-understanding \\
\textit{Participant}      & No, Tokushima.            & 0                & +1               & Reject                   \\
\textit{Robot}         & Oh, Tokushima. I'm sorry. & -1               & 0                & Apology                  \\ \hline
\end{tabular}}
\label{excerpt}
\end{table*}

\begin{table*}[!ht]
\caption{How valence is related to dialogue context -- the robot}
\scalebox{0.91}{
\begin{tabular}{ll}
\hline
\textbf{When is the robot positive}                   & \textbf{When is the robot negative}              \\ \hline
1. In the initial greetings                  & 1. The participant talking excessively \\
2. Introducing itself                             & 2. Talking about the limitations of robots \\
3. The participant saying something amusing & 3. Hearing about the participant’s limitations         \\
4. The participant feeling positive  & 4. The participant showing negative feelings             \\
5. Introducing a new topic                        &                                                   \\
6. Praising the participant                 &                                                   \\
7. The participant answering questions correctly    &                                                   \\
8. Asking a question and expecting a positive answer &                                                   \\ \hline
\end{tabular}}
\label{humanoid}
\end{table*}

\begin{table*}[!ht]
\caption{How valence is related to dialogue context -- the participant}
\scalebox{0.91}{
\begin{tabular}{ll}
\hline
\textbf{When is the participant positive} & \textbf{When is the participant negative}                            \\ \hline
1. In the initial greetings    & 1. Talking about their research                                 \\
2. Talking about their background            & 2. Failing to explain something clearly to the robot        \\
3. Starting a new topic          & 3. Feeling bored with a topic                                 \\
4. Being praised by the robot           & 4. Describing vague topics (e.g., the future, job plans) \\
5. Saying something funny        & 5. Admitting they don’t understand something technical      \\
6. Explaining something vividly  & 6. Being asked a difficult question by the robot          \\
7. Knowing a lot about something (e.g., robots)    & 7. Having to say something negative about the robot       \\
8.  Praising the robot & 8. Being embarrassed or feeling bad for the robot \\
9. Being told their answers are correct by the robot    & 9. Talking about their own limitations   
                                          \\ \hline
\end{tabular}}
\label{participant}
\end{table*}

The valence patterns of the robot and participant are shown in Fig.~\ref{mimicry}. The dialogue can be roughly divided into three phases. We can see that in the spontaneous dialogue phase, the human participant's valence does mimic the robot's to a large extent, especially when the robot changes its valence significantly (e.g., from -2 to +2 and from +1 to -3). Also, during the majority of the time in the spontaneous dialogue phase, the human valence is very close to its previous robot valence, showing a mimicry relationship. Note that, the human participant did not express extremely high valence (i.e., +3). This could be due to individual differences such as cultural background, personality, or expectation of the robot as a novel stimulus. The Pearson’s Correlation Coefficient (PCC) of the valence pairs is 0.54 in the spontaneous dialogue phase, i.e., there is a moderate positive relationship between the human and robot valences. This suggests that it is possible to implement a mapping function between the human's valence and the valence expressed by the SDS. Note that, what we need to investigate is the correlation between emotions for a mapping pattern, not the exact values for classification, so we do not report accuracy or F1 scores. Even if the human participant's emotion values are completely different from the robot's, but highly correlated, e.g., [-3, -2, 1, 0, 1] and [-2, -1, 0, 1, 2], it still shows that the participant's emotion follows the robot, which could be used as a basis for implementing our proposed anticipatory SDS.

Interestingly, during the ice-breaking and ending phases, the human's valence hardly resembles the robot's valence with a PCC of 0.09 in the ice-breaking phase and 0.07 in the ending phase. After examining the video recording, we found that both parties were performing the greeting and leave-taking dialogue acts that they consider to be socially appropriate, instead of mimicking their dialogue partner's expressions. That is, emotions in the dialogue were influenced by DAs. Thus, we annotated DAs following the categorization by \citet{stolcke2000dialogue} to understand how they impact emotion mimicry.

\begin{table*}[!ht]
\caption{Annotation for analyzing the laughter prediction.}
\scalebox{0.91}{
\begin{tabular}{llll}
\hline
\textbf{Speaker} &
  \textbf{Transcript} &
  \textbf{Laughter acoustics} &
  \textbf{Laughter type} \\ \hline
\textit{Participant} &
  \begin{tabular}[c]{@{}l@{}}Although I studied only one night, I passed the\\ exam. \textbf{Haha}.\end{tabular} &
  Flat pitch, moderate power &
   \\
\textit{Robot} &
  \textbf{Hehe}. I see &
    &
  Social laughter \\ \hline
\textit{Participant} &
  \begin{tabular}[c]{@{}l@{}}I was told the exam would be held the following \\ week when I arrived. I had the wrong date. \textbf{Haha}.\end{tabular} &
  Long duration, jittery, shimmery &
   \\
\textit{Robot} &
  \textbf{Ufufufu}. I see. &
   &
  Mirthful laughter \\ \hline
\textit{Participant} &
  \begin{tabular}[c]{@{}l@{}}I studied hard for the exam but got a zero. I was \\ very sad. \textbf{Haha}.\end{tabular} &
  low pitch and power &
   \\
\textit{Robot} &
  That is bad. &
   &
  No laughter \\ \hline
\end{tabular}}
\label{laughteranno}
\end{table*}
 
An annotation excerpt from the ice-breaking phase is shown in Table~\ref{excerpt}. When the robot asked a ``Wh-question'' and the participant responded with a ``statement'', both valence and arousal display mimicry. However, when the robot expressed ``signal-non-understanding'' and the participant responded ``reject'', the valence has an obvious drop, but the arousal barely changes. This shows that unlike valence, arousal is less influenced by DAs.

In terms of arousal, we found significant mimicry in the participant's arousal toward the robot's (figure omitted for brevity). The PCC of the arousal pairs is 0.78 over the whole dialogue session, including the ice-breaking and ending phases. This demonstrates that the arousal behind the future response may be relatively easy to predict based on the current utterance.

Our preliminary analysis indicates that it is feasible to predict a future emotion from the current emotion and DA for implementing the prediction component of our proposed architecture. We also found that valence is related to contextual information, such as DAs and personal factors of the interlocutor. We summarized a set of observations on the relationship between dialogue context and the valence of the robot and of the human participant in Table~\ref{humanoid} and Table~\ref{participant}, respectively. We expect that these observations can contribute to the future implementation of the proposed anticipatory SDS and to the broader research community.

\subsection{Exploring the Feasibility of Laughter Prediction}

We analyzed a publicized dialogue demo that was built upon the second corpus with the robot's dialogue system replacing the teleoperator\footnote{\url{https://www.youtube.com/watch?v=6tMiWog4l00}}.
The results are presented in Table~\ref{laughteranno}. As shown here, the robot generated suitable laughter behaviors based on the acoustic features of the previous user laughter. When the user's laughter had a flat pitch and moderate power, the robot responded with social laughter. When the user's laughter had a long duration and was jittery and shimmery, the robot responded with mirthful laughter. The robot also ``understood'' not to laugh when the user laughed only to relieve embarrassment. Based on previous research on the contagious phenomenon of laughter \cite{provine1992contagious}, we aim to expand on the shared laughter research by adjusting the acoustic features of the laughter generated by the SDS, allowing it to predict the future laughter behaviors of the user in response to the system's laughter.

\section{Discussion}
Our preliminary analysis of human-robot dialogue suggests that it is feasible to implement an anticipatory SDS that predicts the emotion and laughter of the user in the next turn using the current turn of the system. However, there remain open challenges in implementing the proposed architecture, which we will discuss in this section. Further, we will discuss potential application scenarios for the proposed architecture.

\subsection{Implementation Challenges}
The proposed architecture relies on accurate and robust recognition of emotions from speech, which remains an open challenge due to the variability in speech and emotion expression. The prediction component in the proposed architecture may not be applicable to all user turns, as humans can express emotions and laughter arbitrarily without considering the system's expression. In this case, the architecture can be ``downgraded'' with only emotion recognition and laughter detection working and the prediction and self-correction components frozen. Further, noise in real-world applications of SDSs can reduce the reliability of both the emotion recognition and laughter detection models. Therefore, we plan to incorporate other communicative modalities (e.g., text and vision) to further improve the proposed architecture \cite{li2020attention}.

\subsection{Potential Application Scenarios}
In social and open-domain dialogue, an SDS with our proposed architecture can generate appropriate conversational and affective behaviors, such as a backchannel ``Yeah'' or laughter, in real time, instead of having delayed turn-taking that may interrupt the user's next turn. Further, it is especially useful in scenarios where the SDS is expected to take initiatives and lead the conversation, such as in healthcare and education. For example, the SDS can adjust its generated emotions and DAs to support a user's emotion regulation process by eliciting certain emotions in people with depression or autism \cite{lubis2018eliciting}. In education, the SDS can express emotions during collaborative problem solving with children to increase their participation in the learning activities and resulting learning outcomes, as \citet{zhou2020would} found that when the robots exhibited emotional expressions, participants were more likely to collaborate with them and achieve task success faster.

\section{Conclusions}
In this work, we propose an anticipatory SDS architecture that predicts the affective reactions of the user in a future turn using its behaviors in the current turn. We investigate its viability in both speech and laughter scenarios. Based on preliminary analysis of human-robot dialogue, we demonstrated that: 1) The emotion of a future turn can be predicted from the current turn. The arousal dimension has a significant mimicry relationship, in which the human user's arousal follows the robot's arousal during dialogue. The valence, however, is also related to the previous DA. 2) The laughter behavior of the human user in a future turn has a mapping pattern with the laughter behavior of the robot in the current turn. The preliminary analysis paves the way for our future research. In particular, we aim to identify the relationship between current DA and future emotion, as well as current laughter and future laughter to implement the emotion prediction model and the laughter prediction model of the proposed architecture. Moreover, we plan to include history information and dialogue context beyond the current turn to improve the prediction accuracy. Achieving anticipatory SDSs requires every individual component to be accurate and robust, as well as a seamless collaboration between the components. Thus, in the future, we plan to implement the complete architecture and evaluate its outcomes in user studies.


\balance
\bibliographystyle{ACM-Reference-Format}
\bibliography{sample-base}




\end{document}